\documentclass[sigplan]{acmart}
\usepackage{subcaption}
\usepackage{booktabs}
\usepackage{physics}
\usepackage{xcolor}
\usepackage{microtype}
\AtBeginDocument{%
  }

\acmConference[arXiv preprint]{arXiv preprint}{April}{2026}
\acmBooktitle{Architecting Early Fault Tolerant Neutral Atoms Systems with Quantum Advantage}
\acmPrice{}
\acmISBN{}
\acmDOI{}

\setcopyright{none}
\renewcommand\footnotetextcopyrightpermission[1]{}
\begin{document}

\title{Architecting Early Fault Tolerant Neutral Atoms Systems with Quantum Advantage}

\author{Sahil Khan}
\email{sahil.khan@duke.edu}
\orcid{0009-0000-4160-8010}
\affiliation{%
  \institution{Duke University}
  \city{Durham}
  \state{North Carolina}
  \country{USA}
}

\author{Sayam Sethi}
\orcid{0009-0005-3056-5285}
\affiliation{%
  \institution{University of Texas at Austin}
  \city{Austin}
  \state{Texas}
  \country{USA}
}

\author{Kaavya Sahay}
\affiliation{%
  \institution{Yale University}
  \city{New Haven}
  \state{Connecticut}
  \country{USA}
}

\author{Yingjia Lin}
\affiliation{%
  \institution{Duke University}
  \city{Durham}
  \state{North Carolina}
  \country{USA}
}

\author{Jude Alnas}
\affiliation{%
  \institution{Duke University}
  \city{Durham}
  \state{North Carolina}
  \country{USA}
}

\author{Suhas Kurapati}
\affiliation{%
  \institution{Duke University}
  \city{Durham}
  \state{North Carolina}
  \country{USA}
}

\author{Abhinav Anand}
\orcid{0000-0002-8081-2310}
\affiliation{%
  \institution{Duke University}
  \city{Durham}
  \state{North Carolina}
  \country{USA}
}

\author{Jonathan M. Baker}
\orcid{0000-0002-0775-8274}
\affiliation{%
  \institution{University of Texas at Austin}
  \city{Austin}
  \state{Texas}
  \country{USA}
}

\author{Kenneth R. Brown}
\orcid{0000-0001-7716-1425}
\affiliation{%
  \institution{Duke University}
  \city{Durham}
  \state{North Carolina}
  \country{USA}
}

\begin{abstract}
  Recent advancements in neutral atom platforms have enabled exploration of early fault-tolerant (FT) architectures for applications with quantum advantage, such as quantum dynamics simulations. An efficient fault-tolerant architecture has both spatially efficient quantum error correction codes (low qubit overhead), and efficient methodologies (transversal based gates, extractor based gates, etc.) for logical computation, to minimize overall execution time. Achieving the right balance between space and time can be critical for enabling early FT demonstrations of quantum advantage. 

In this work, we identify bottlenecks in existing spatially efficient schemes, which tend to be very serial, and do not take advantage of unutilized space. We introduce a teleportation-based scheme that leverages the reconfigurable connectivity of neutral atoms to parallelize logical operations. Our approach achieves up to \textbf{$\mathbf{\sim 3 \times}$ speedup} over extractor architectures at no extra space cost and achieves the best spacetime performance among other viable architectures before accounting for external \textit{resource-states}. To rigorously evaluate performance, we construct explicit quantum advantage benchmarks and \textit{simulate} compilation to a fault-tolerant instruction set, including low-level gate scheduling and shuttling patterns, and resource-state nondeterminism. We find that our speedups still apply and report exact space-time cost along with success probabilities, identifying architectures capable of achieving quantum advantage \textbf{with as little as $\mathbf{11,495}$ atoms and a runtime of $\mathbf{\sim 15}$ hours}.
\end{abstract}


\renewcommand{\shortauthors}{Khan et al.}

\maketitle

\section{Introduction}

Quantum computers promise a computational advantage over classical computers in  applications across scientific simulations, optimization, and factoring problems ~\cite{shor1999polynomial, kivlichan2020qsimelectrons, childs2018firstqsim, harrow2009hhl, reiher2017nitrogenfixation, quantum_algos_table, early_ft_fermi_hubbard, analog_sim}. The qubits used to perform computation are highly susceptible to errors; therefore, quantum error correction (QEC) must be implemented to ensure successful program execution. QEC codes use many noisy physical qubits to encode information into reliable \textit{logical} qubits which can be used to construct large-scale fault-tolerant architectures. For a given code, logical programs must be compiled into a native gate-set supported by the code, and this combined choice of code and logical compilation method constitutes the fault-tolerant architecture.

\subsubsection*{Architecting on Current Neutral Atoms Systems}

Recent developments in neutral atoms hardware have demonstrated large numbers of atoms~\cite{6100qubits, logicalqubitsneutrals_atom_computing, pasqal_1000_atoms_2024, atom_computing_ac1000, logical_msd_on_neutral_atom_demonstration}, with some experiments controlling up to 6100 coherent atoms~\cite{6100qubits}. In addition, there have been demonstrations of low physical error rates and experimentally realized QEC ~\cite{logicalqubitsneutrals_atom_computing, logical_msd_on_neutral_atom_demonstration, ftarchitectureNA}. These advances allow for QEC codes with sufficient distance to maintain low overall application error rates under realistic hardware noise models and to explore different designs for early fault-tolerant architectures. Such explorations aim to identify which designs are most practical in terms of both space (physical qubit footprint) and overall execution wall time. For at least the near-term and smaller quantum advantage applications, particularly dynamics simulations, a proper balance of resource tradeoffs can be the difference between enabling practical execution of quantum advantage or not. Crucially, the measurement times in neutral atom systems are approximately $\sim 1000\times$ slower than physical gate times, accounting for nearly all of the fault tolerant application's execution. This slow measurement time gives motivation to find parallel and spacetime-efficient architectures on neutral atoms that parallelize logical operations and thus their respective measurements.

\begin{figure*}[htbp!]
    \centering
    \includegraphics[width=0.95\linewidth]{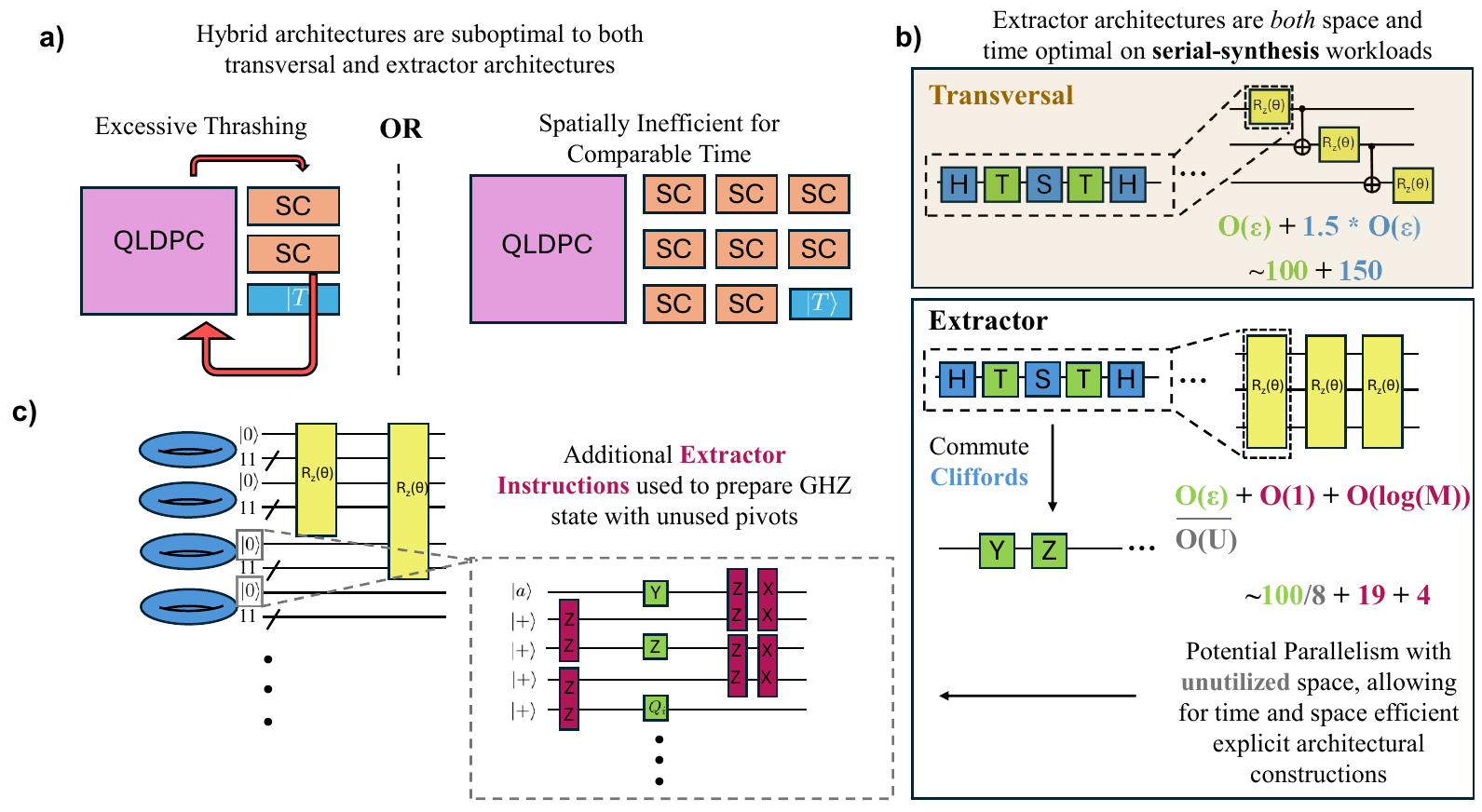}
    \caption{a) Hybrid architectures with low spatial requirements suffer from severe thrashing, where the load/store times increase wall time beyond the space efficient extractor. Once the equivalent time bounds are met, these architectures are too spatially demanding. 
    b) A transversal architecture performs poorly in terms of time and space on serial-synthesis workloads due to the extra Clifford operations (blue) and serial non-Cliffords (green), whereas an extractor architecture performs well in terms of time even with the additional extractor instructions (raspberry color) at the $R(\varphi)$ level. We give the asymptotic formulas of the time cost per $R(\varphi)$ rotation in terms of precision, $\epsilon$, number of modules, $M$, and unutilized modules $U$. We then give a sample realistic case within the range of practical assumptions we detail in this paper below it. We detail a scheme to further parallelize the non-Cliffords using no additional space in c) where we take advantage of the relaxed connectivity constraints of neutral atoms and the \textit{unutilized} modules to speed up computation using a teleportation-based scheme.
    }
    \label{fig:fig1}
\end{figure*}

\subsubsection*{Parallelizing Extractor Architectures}
In this work, we are motivated to identify the most practical architectures for executing the first dynamics simulations with quantum advantage on neutral atoms systems. We compare the spatial and temporal overheads of \textit{three} different classes of candidate architectures: \textit{Transversal} operations which realize a logical gate through the combination of physical gates between two code blocks in a fault tolerant manner, \textit{Extractor} based operations (also known as Code Surgery-based) which rely on projective measurements between two code blocks, or a \textit{Hybrid} approach which loads and stores between the two \cite{constantoverheadqecneutralatoms, HetEC, Viszlai_generalized_bb}. The largest source of slowdown on extractor architectures, which are the most spatially efficient, comes from the large number of serial operations in a subroutine known as \textit{$R(\varphi)$ synthesis}. Extractor architectures contain \textit{unutilized} space during this synthesis stage. Furthermore, connectivity constraints are much more relaxed in neutral atoms, allowing for parallelization schemes to interact with this unutilized space at little cost. We propose a gate-teleportation-based scheme to make use of the previously unutilized extractor modules to accelerate computation. The speedup scales as $O(U)$, where $U$ is the number of unutilized modules, whereas transversal-based parallelization is limited to a constant rate for a given application (without incurring additional space). We find that our scheme allows spatially optimal constructions to also be temporally optimal in certain dynamics applications, improving over transversal architectures in terms of \textit{both} space and time. We find two criteria that enable \textit{both} our spatially efficient extractor scheme and base extractor architectures to be competitive with transversal architectures. Namely, 1) $R(\varphi)$ synthesis comprises the majority of the circuit, which is typical in most quantum algorithms that are not classically simulable, and 2) The transversal circuit cannot parallelize more than $C\%$ of the synthesis cost, where $C$ is the ratio of Clifford to T gates in $R(\varphi)$ synthesis.

\subsubsection*{Architectural Simulations of Quantum Advantage}
Speedup can be more accurately studied by accounting for full compilation to a fault-tolerant instruction set, low-level gate and shuttling patterns, and T-state factory nondeterminism. This provides a more precise assessment of application success probability and wall time than prior works based solely on resource estimation \cite{cain2026shorsalgorithmpossible10000, prakash-beverland-resource-estimates, quantum_algos_table, tourdegross}. Our parallelization scheme's performance gains are noticeable even when realistic T-state throughput is achieved in early fault-tolerant demonstrations, as detailed in Tables \ref{tab:heisenberg} -- \ref{tab:fermi}. Our explicit constructions using the two-gross code combined with our parallel injection scheme achieve better than existing performance in both space and time for these applications compared to other existing architectures, supported by an adequate number of special \textit{resource state} factories (such that our scheme's wins are not diluted by starvation in other areas). We find architectures that can achieve quantum advantage with as few as \textbf{11,495 atoms and $\sim$ 15 hours} for certain dynamics simulations, which we believe to be a realistic goal for the near future.

Our contributions are as follows: \begin{enumerate}
    \item We identify the main source of slowdown in extractor architectures as serialization at the injection level and propose a scheme that takes advantage of \textit{unutilized} space. This yields up to $\sim3\times$ speedup over the base extractor without additional space cost, and up to $17\%$ speedup over transversal architectures such that even if high rate transversal and addressable gate sets were to be found with equivalent space, our proposed architectures are still ideal. 
    \item We outline explicit constructions of an extractor architecture with a complementary cultivation scheme capable of performing four key fault tolerant dynamic simulations with quantum advantage through realistic simulations of device error rates. We replace previous analytical estimates with simulation of extractor architectures using a defined compute model, and find that our architecture globally minimizes spacetime and is best in both space and time for 2 of our 4 quantum advantage benchmarks.
    \item We replace previous resource estimates with more precise simulation of the time/error overhead of each compiled instruction, resource state nondeterminism, and low-level interactions. We find that our results hold through these simulations, as we still achieve the best space and time for 2 of the 4 quantum advantage benchmarks, and are competitive in spacetime in all 4.
    \item We detail criteria for when extractor architectures can outperform a transversal gate architecture, and find that hybrid load/store architectures are not sufficient in space nor time for early demonstrations of quantum advantage on neutral atoms.   
\end{enumerate}

\section{Background}

\subsection{Quantum Error Correction}
Practical quantum systems suffer from noise, which affects the fidelity and success rate of the quantum programs being executed on these systems. Various quantum error mitigation techniques have been proposed in the NISQ era, however, their utility decreases as system size scales. Quantum Error Correction (QEC) on the other hand, enables exponential suppression of the logical qubit error rates as the distance of the code increases. A Quantum Error Correcting Code (QECC) with code parameters $[[n, k, d]]$ encodes $k$ logical qubits into $n$ physical qubits, with a code distance $d$. The distance $d$ indicates how many independent physical qubit errors can cause a logical error on the code. A QEC code with distance $d$ can detect up to $d - 1$ errors and can successfully correct up to $\lfloor(d - 1) / 2\rfloor$ independent errors. The encoding rate of a QEC code, which is denoted by $k/n$ is another metric that indicates the \textit{efficiency} of the code. Codes with higher encoding rates are preferred as they are more spatially efficient, thereby alleviating the physical qubit resource requirement.

\subsubsection{Bivariate Bicycle Family}
A QECC family that has recently gained popularity is the Bivariate Bicycle (BB) Code family, specifically the Gross and Two Gross Codes~\cite{bbcodes, tourdegross}, with code parameters $[[144, 12, 12]]$, and $[[288, 12, 18]]$, respectively. 
These codes also have a high encoding rate, which makes them promising for both near-term and large-scale fault-tolerant quantum computation (FTQC).
Logical quantum computation on both these code instances has been well studied~\cite{he2025extractorsqldpcarchitecturesefficient, tourdegross}. 
We place special emphasis on these codes and use them as the default QLDPC code due to their appeal to industry and well-defined compute model~\cite{ibm_blog_ftqc_2025, tourdegross}. However, our results generalize to other QLDPC codes, as illustrated in Figure \ref{fig:sensitivity-codes}.

\begin{figure}[htbp!]
    \centering
    \includegraphics[width=0.99\linewidth]{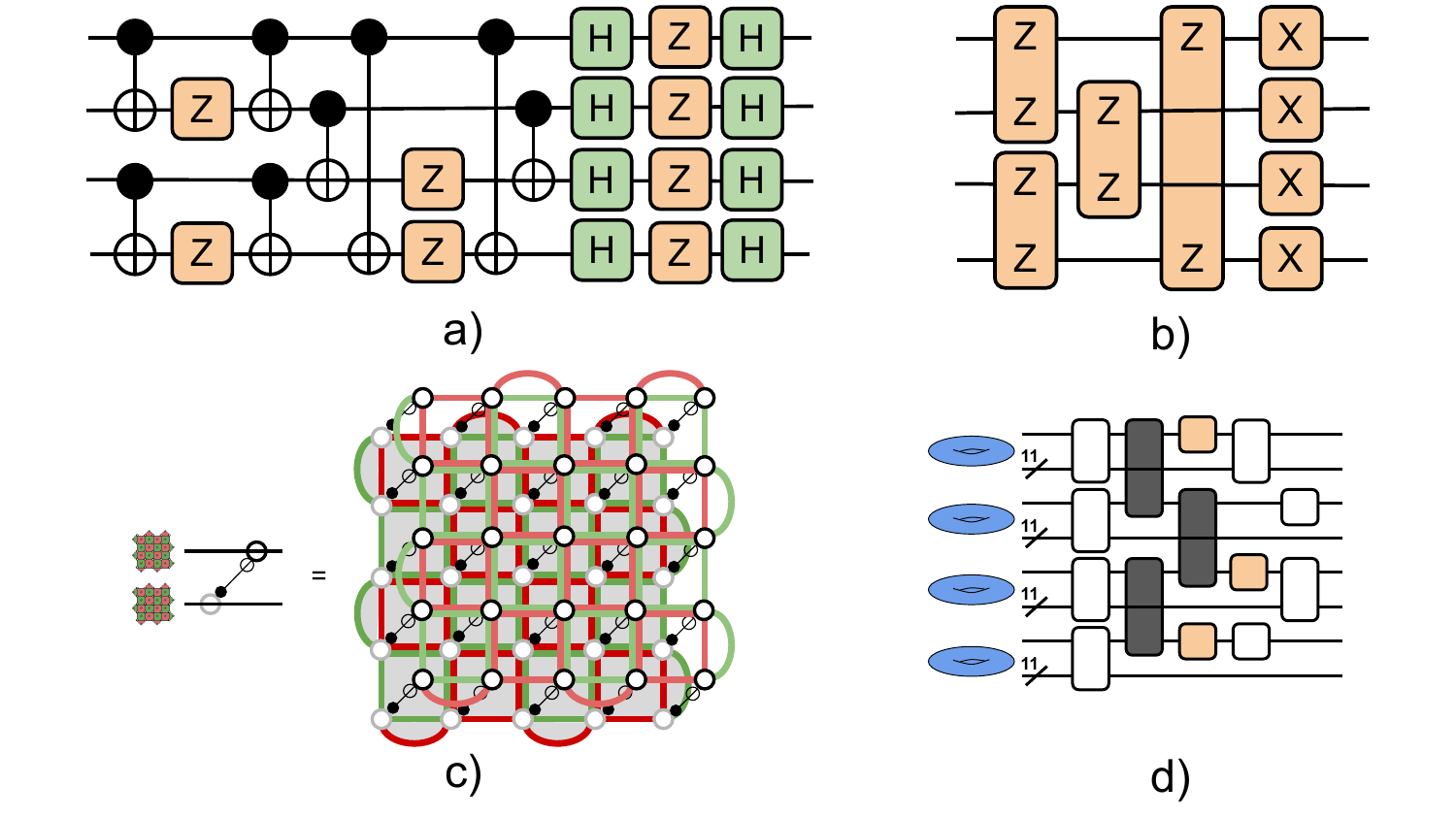}
    \caption{(a) Illustration of the quantum circuit considered in this work. (b) Corresponding compiled Pauli-based computing (PBC) instruction. (c) A transversal CNOT gate between two logical qubits encoded in the surface code that can be used to implement the circuit shown in (a). (d) Compiled PBC instruction (b) mapped onto the gross code architecture. For further details on the compilation procedure to transversal gates and Pauli-based computing, see Ref.~\cite{litinski2019gameofsurfacecodes, moussa2016transversal, kubica2015universal}}
    \label{fig:pbc_trans}
\end{figure}

\subsection{Compilation Methodologies}

We describe three different quantum circuit compilation strategies, illustrated in Figures~\ref{fig:pbc_trans} and \ref{fig:hybrid-archs-explain}.

\subsubsection{Transversal-Based Compilation}
Transversal-based compilation relies on a QEC code, here, the surface code, that admits transversal logical gates. In this approach, each application-level gate is mapped to a logical operation implemented transversally across code blocks. A transversal single-qubit logical gate applies the corresponding physical gate independently to each qubit within a code block, while a transversal two-qubit logical gate applies pairwise physical gates between corresponding qubits in separate code blocks.. A sample program and its transversal implementation of a CX is shown in Figures \ref{fig:pbc_trans}a and \ref{fig:pbc_trans}c, respectively.

\subsubsection{Extractor-Based Compilation}
\label{sec:extractor-background}
Extractor-Based Architectures must first compile to the target application from the gate level instruction sequence to a Pauli-Based Computing (PBC)~\cite{tourdegross} representation. PBC is a paradigm of quantum computation where every operation is either a resource state preparation, or a multi-qubit projective measurement (Figure \ref{fig:pbc_trans}b). Compilation to the PBC model has been widely studied~\cite{litinski2019gameofsurfacecodes,litinski_lattice_2018,tourdegross}, and is especially relevant for newly discovered QEC codes since it enables universal quantum computation via an extractor based architecture~\cite{he2025extractorsqldpcarchitecturesefficient}. The input program is first compiled into a PBC circuit (Figure~\ref{fig:pbc_trans}a to Figure~\ref{fig:pbc_trans}b), where each gate corresponds to a projective measurement. The logical qubits involved in the circuit are then mapped to different modules on the device~\cite{sethi_optimizing_2026}. Operations between logical qubits within the same module are performed via in-module measurements, whereas the operations between logical qubits on different modules are performed via inter-module measurements. Every projective measurement in the original PBC circuit is broken down into a sequence of in-module and inter-module measurements that are supported by the quantum system's Instruction Set Architecture (ISA), as shown in Figure~\ref{fig:pbc_trans}d. For example, a Bivariate Bicycle ISA~\cite{tourdegross} consists of only five instructions: idling (syndrome extraction), shift automorphisms (permutations of code stabilizers to create some Cliffords), in-module measurements (which combine with shift automorphisms to create any arbitrary Clifford), inter module measurements, which allow logical connectivity between modules, and T injections. \textbf{To our knowledge, the Bivariate Bicycle ISA is the only well-defined extractor instruction set} with an exact compilation methodology and explicit cost modeling of instruction execution times and error rates.

\subsubsection{Hybrid (Load-Store) Compilation}
Hybrid or load/store architectures~\cite{ldstr, Viszlai_generalized_bb, HetEC, constantoverheadqecneutralatoms} combine two or more codes as memory, compute, and load/store qubits between the memory section of the computer and logical compute area. We emphasize that these hybrid architectures should not be confused with hybrid hardware platform architectures, which involve physically distinct qubit technologies. In this paper, we evaluate hybrid architectures as one type of high rate QLDPC code combined with one type of transversal-admitting code.  

\begin{figure}[htbp!]
    \centering
    \includegraphics[width=0.99\linewidth]{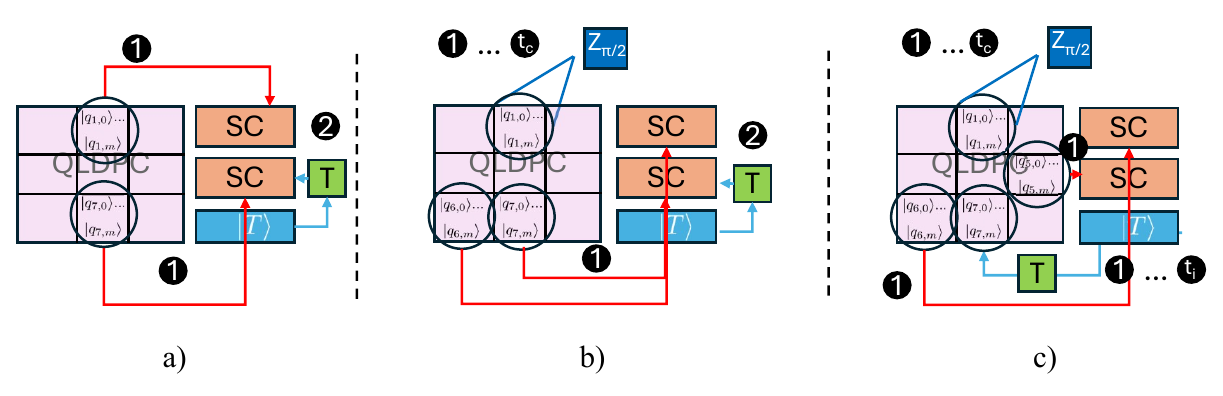}
    \caption{(a) A simple hybrid compilation policy that delegates all computation to the surface code. This policy performs a compute region store in parallel in step 1, as well as performing a T-state injection after loading in step 2. (b) A hybrid compilation policy that builds off (a) and allows for in-memory compute on single qubit gates of modules, gaining additional parallelism but taking the entire module offline for $t_c$ steps (c) a compilation policy with the same assumptions of (b), with the added capability to perform T-state injections directly onto the QLDPC code, taking a module offline for $t_i$ steps}
    \label{fig:hybrid-archs-explain}
\end{figure}

These architectures allow for the ease of implementation in transversal gates while maintaining control on qubit overhead. A hybrid architecture allows for the transfer of logical information through QLDPC code surgery, where a ZZ projective measurement and a controlled X measurement allow for a store or load operation within two $O(d)$ measurement timesteps. Logical computation is then performed strictly on the transversal-admitting code unless otherwise specified. Computation thus proceeds layer by layer for the logical circuit with many loads/stores mixed in depending on the capacity of the compute region, similar to the cache eviction problem in the classical computing sense. Prior work on hybrid architectures have used an earliest instruction first policy, greedily choosing the most locally optimal choice considering the next layers' instructions and evicting the qubits used furthest in future \cite{Viszlai_generalized_bb}. Other work has since attempted to make use of the fact that "in-memory" compute is possible with extractor based architectures, but made simplifying assumptions about precise cost-modeling. We extend this work and model the compilation policy among other sample hybrid compilers as shown in Figure \ref{fig:hybrid-archs-explain}.

\subsection{Neutral Atoms Systems}
\label{sec:background-na}
\subsubsection{Basic Constraints and Measurements}
Neutral atom quantum computing uses reconfigurable optical tweezer arrays for trapping, moving, and addressing individual qubits. A physical circuit can be implemented using \textit{low-level operations}: physical gates, shuttling (movement operations), and measurements. A wide range of shuttling and gate times have been demonstrated \cite{saffman_neutral_atoms_parameters, neutral_atoms_gate_times_1, dolev_gate_and_shuttle_times, Levine_2019, survey_neutral_atoms_timings}, typically on the order of microseconds. Measurement times, however, are notably much slower than all of the other low-level operations, on the order of milliseconds \cite{survey_neutral_atoms_timings, measurement_times_readout_Xu_2021, 5ms_readout}. This leads to execution dependent on QEC cycles in the milliseconds, due to fault-tolerant circuits requiring mid-circuit measurement. \textbf{For this reason, fault tolerant neutral atoms compilers must focus on reducing the amount of required measurement steps}. An exact table of all of our assumed hardware parameters is shown in Table \ref{tab:params}.

\subsubsection{Allowable Movement Patterns and Gate Operations}
Neutral atom arrays can support long-range interactions, yet these restrict parallelism due to their larger Rydberg blockade radius. Long-range gates are confined to a maximum interaction distance. Atoms can be moved around through shuttling, however, if all necessary interactions satisfy the maximum interaction distance, it can be faster to serialize all gates than to shuttle atoms and perform gates in parallel as seen in \cite{fast_bb}. This is due to the fact that gates are on the order of a few microseconds, whereas shuttling to even a neighboring spot can be an order of magnitude slower. Atoms are allowed to move subject to the constraints of crossed acoustic optic deflectors (AODs), where a set independent horizontal and set vertical component fixes the translation for atoms to move in parallel across arbitrary amounts of atoms in a single time step \cite{dolev_gate_and_shuttle_times, constantoverheadqecneutralatoms}. This allows for parallel movement and gates across the entire machine so long as the set of movement and gate patterns for a single timestep is \textit{translationally symmetric}.

\subsubsection{Realtime Decoding and Classical Corrections}
Fault tolerant quantum computation requires classical feedback loops. These include the decoding of error syndromes and applying the corresponding error correction, or tracking whether T injections were successful and fixing the corresponding Pauli frames accordingly \cite{litinski2019gameofsurfacecodes, real_time_decoding_holistic_review, relay_bp, vittal2023astrea, fastforcurious}. Long measurement times in the milliseconds allow for the effects of classical feedback loops to be diluted, just as gates times are, as these take on the order of a microsecond. While real-time decoding for QLDPC codes in the nanosecond scale is an open problem for superconducting systems, for neutral atoms this is not a problem. For the rest of this paper, we assume decoding as well as classical feedback for Pauli-frame-tracking can be done on the microsecond scale while maintaining performance~\cite{turnerscalabledecoding}.

\subsection{Quantum Advantage Benchmarks}
\label{sec:background-qa}
Simulating the dynamics of quantum spin models is of interest in quantum information science, condensed matter physics, and chemistry.
Exact classical simulation of such quantum systems require computational resources that grow exponentially with system size.
Although approximate methods such as quantum Monte Carlo and tensor network techniques can accurately compute some ground-state and static properties, they are limited for real-time dynamics, especially in strongly correlated regimes.
For example, it was shown in Refs.~\cite{flannigan2022propagation, daley2022practical} that simulating the time dynamics of the 2D transverse-field Ising model (TFIM) and the Fermi–Hubbard (FH) model on a $10\times10$ lattice is beyond the capability of current tensor network methods.
Thus, in this paper, we consider systems of similar scale that are beyond the reach of state-of-the-art classical simulation methods.

In particular, we consider four systems: the Heisenberg model~\cite{anderson1959new, flannigan2022propagation, yoshioka2024hunting}, the nearest-neighbour (NN) and long-range (LR) transverse-field Ising models~\cite{cipra1987introduction, flannigan2022propagation}, and the Fermi-Hubbard~\cite{hubbard1963electron} model.
For the Fermi-Hubbard model, we use the Jordan–Wigner transformation~\cite{jordan1928paulische} to map fermionic operators to qubit operators.
We simulate the time dynamics by approximating the time-evolution operator, $e^{-i\hat{H}t}$, using higher-order Trotter–Suzuki product formulas~\cite{suzuki1976generalized}.
We construct efficient circuits for time evolution by grouping the Hamiltonian into sets of commuting terms and constructing Clifford circuits for simultaneous diagonalization, following the techniques in Refs.~\cite{anand2025hamiltonian, anand2025leveraging, van2020circuit}. 
The resulting diagonal unitaries are then scheduled in parallel using a greedy algorithm to minimize circuit depth and maximize parallelization.
A detailed summary of system parameters, circuit constructions, and resource estimates is provided in Table~\ref{tab:ft_summary}.
We refer the reader to ~\cite{quantum_algos_table} for a detailed complexity analysis. For all benchmarks except the 2D Heisenberg model, 1 Trotter step is assumed as sufficient precision for simulation. For the 2D Heisenberg model, we find 300 trotter steps is necessary for practical precision \cite{childs2018firstqsim} at the sixth order Trotter approximation of this benchmark (obtained using a curve fit). This makes the benchmark large and computationally demanding due to the immense classical resources needed. Therefore, we run simulations using a single trotter step and subsequently scale the resulting time and error estimates accordingly.

\begin{table}[htbp!]
\centering
\scriptsize
\setlength\tabcolsep{0pt}
\caption{Summary of fault-tolerant resource estimates for simulating the dynamics of different models. For all applications, we set the interaction strengths $J$, $J_x$, $J_y$, $J_z$ = 1, transverse field $B$ = 1, power law decay coefficient $\alpha$ = 2, hopping term $t$ = 1, onsite interaction strength $U$ = 1, and synthesis precision $\epsilon$ = $10^{-10}$. The trotter steps $\delta$, and evolution time $T$, which may be a function of the logical qubits $N$, are variable parameters}
\label{tab:ft_summary}
\begin{tabular*}{\columnwidth}{@{\extracolsep{\fill}}lccccc}
\toprule
Model & Method & Lattice & \#T Gates & $N$ & Parameters \\
\midrule
2D Heisenberg & Trotter (Sixth Order) & (5$\times$10) & $1.5 \times 10^{7}$ & 50 &  $\delta$ = 300, T = $N$ \\ 
2D LR TFIM & Trotter (Fourth Order) & (10$\times$10) & $5 \times 10^{6}$  & 100 & $\delta$ = 1, $T = 10$ \\
2D NN TFIM & Trotter (Fourth Order) & (10$\times$10) & $3 \times 10^{5}$ & 100 & $\delta$ = 1, $T = 10$ \\
Fermi--Hubbard  & Trotter(Fourth Order) & (10$\times$10) & $1.1 \times 10^6$ & 200 &  $\delta$ = 1, $T = 1$ \\
\bottomrule
\end{tabular*}
\end{table}
\section{Parallelizing Extractor Based Architectures}
\label{sec:parallelize-injection}
In this section, we first break down the cost  of base extractor architectures, and identify the dominant source of time overhead: serial non-Cliffords. We make the key insight that the flexible connectivity of neutral atom platforms enables the use of previously unutilized modules to accelerate computation at no extra space cost. First, we make the trivial optimization for our scheme that the architecture can parallelize at the $R(\varphi)$ level, such that if there are commuting $R(\varphi)$ rotations, they can be performed in parallel with an identified injection pivot chosen arbitrarily for each rotation so long as there are enough factories to support them. We then base our teleportation-based parallel injection scheme on two observations, and detail our algorithm.

\begin{figure}[htbp!]
    \centering
    \includegraphics[width=\linewidth]{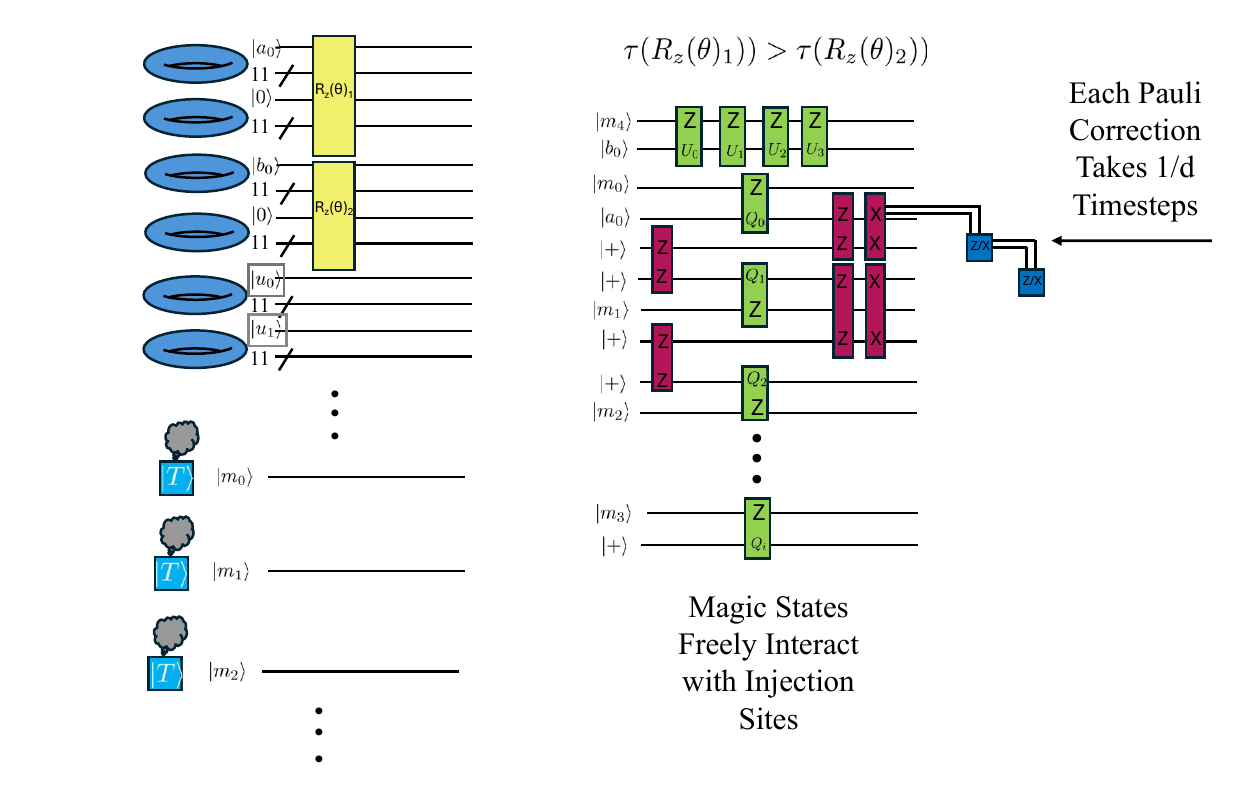}
    \caption{We detail our teleportation scheme to parallelize T-state injection for the largest synthesis length at a time, found by cost $\tau(Rz(\theta)_i)$. On the left, we show the circuit where a given layer consists of $R(\varphi)$ rotations, and an appropriate injection pivot identified arbitrarily once per each $R(\varphi)$, we label $a_0$, and $b_0$. The T states are created by $O(M)$ T factories with magic states labeled $\ket{m_0}$. The parallelization scheme then takes the largest synthesis length and parallelizes the largest synthesis across unutilized pivots (gray) while the other $R(\varphi)$ rotations and corresponding injection pivots proceed serially. We show how this requires a separate factory per T injection site, and how each Clifford correction takes one destructive measurement as opposed to one timestep unit ($d$ measurement rounds) for projective measurement, therefore allowing for parallelization into the next layer.}
    \label{fig:teleportation-scheme}
\end{figure}

\subsection{Analyzing Time Overhead in Extractor-Based Architectures}
Extractor-only architectures perform computation entirely on a high-rate code, achieving minimal spatial overhead. We define one timestep as $d$ rounds of measurement. This choice is motivated by the observation that measurement operations are slow on neutral atoms, and $\sim1000 \times$ more costly than the other low-level instructions. Even in shuttling or gate intensive operations, measurement accounts for the majority of the total operation latency. This is further analyzed in Section \ref{sec:evaluation} and explicitly modeled in Section \ref{sec:shuttling-and-gate}. 

We compile using the well-defined instruction cost model of the bivariate-bicycle extractor architecture from Ref.~\cite{tourdegross}. As in Ref.~\cite{tourdegross} we compile applications into $R(\varphi)$ rotations, as this reduces entanglement overhead per rotation, and allows for greater parallelism across PBC rotations. For a rotation acting on $M$ modules, extractor architectures require $O(log_2(M))$ timesteps to distribute entanglement to all participating modules. This follows from the observation that entanglement can be distributed in a binary-tree fashion where the number of entangled modules doubles after each measurement round. 

The subsequent \textit{in-module measurement synthesis} can be performed in parallel across modules, however, performing in-module measurements can be costly. For Bivariate Bicycle codes, this in-module measurement cost is between 1 and 24 $d$ measurement rounds, where $d$ is the code distance, with an average of 18.5, where the total cost incurred for these instructions is the maximum cost among all modules. Crucially, \textit{extractor instructions} are inserted once per $R(\varphi)$ rotation. Each $R(\varphi)$ synthesis requires $O(log(1/\epsilon))$ T gates, which for all practical values of $\epsilon$ greatly exceeds the constant number of extractor instructions. For example, at our chosen practical value of $\epsilon$, $10^{-10}$ (we study sensitivity to this value later in figure \ref{fig:sensitivity-codes}), the majority of the time cost arises from $R(\varphi)$ synthesis itself. This leads us to two observations: 
\begin{figure}[htbp!]
    \centering
    \includegraphics[width=0.85\linewidth]{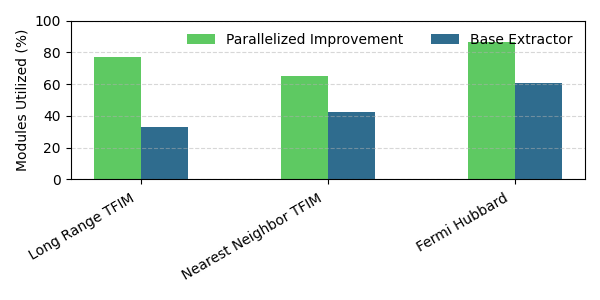}
    \caption{Across the three QA benchmarks where our parallel injection is allowed to parallelize, we see an increase in the number of modules utilized}
    \label{fig:modules_utilized}
\end{figure}

\subsubsection{Observation 1: R($\varphi$) synthesis is the Main Bottleneck}

 For our base assumption of $\epsilon = 10^{-10}$, synthesis takes nearly $\sim$ 100 T gates, translating to 100 non-Clifford injections and thus 100 timesteps. Meanwhile the maximum number of extractor instructions is 24 \cite{tourdegross, pinnacle-architecture} for in-module measurements and $log_2(M)$ for inter-module measurements where $M$ is the number of modules. Although the $log_2(M)$ terms appears to scale unfavorably asymptotically, the required number of necessary space for an application's inter-module time to be equivalent to the synthesization process is $2^{100}$ modules $\times$ qubits per module. All three steps can be parallelized and executed independently, meaning the maximum is the only factor that effects total wall time. 

\subsubsection{Observation 2: Many Modules are Unused}

The non-commuting R($\varphi$) rotations leave spaces within a circuit where entire modules remain idle while waiting on the rest of computation. Furthermore, each module contains a logical ancilla at the pivot which is reset after each round. For this reason, we devise a scheme that can utilize this logical ancilla to create a GHZ state similar to \cite{litinski2019gameofsurfacecodes} and use gate teleportation to execute anti-commuting gates in parallel.

\subsection{Parallelization Circuit and Algorithm} 
To implement this teleportation-based scheme, we divide the parallelization gadget into three stages: preparation, execution, and teardown. 
In the preparation stage, we perform ZZ measurements between pairs of freshly reset logical ancillas initialized in the $\ket{+}$ state. For each parallel layer, two additional ancillas are entangled via ZZ measurement. The entire preparation stage requires a single timestep. 
In the execution stage, we perform T-gate injection using the magic state factories. Each non-Clifford injection requires one timestep.
In the teardown stage, ZZ and XX measurements are performed on the ancillas, followed by Pauli frame corrections. This disentangles the temporary resources and restores the standard configuration. The teardown stage requires two timesteps. The degree of parallelization is therefore limited by the number of unutilized modules.

Precisely, the amount of parallel gates that can be executed in a single timestep is $\frac{u}{2} + 1$, where $u$ is the number of unutilized ancillas. Since preparation and teardown together require three timesteps, we only apply this optimization when it can outperform four serial steps (i.e., when at eight unutilized modules are available). However, we note that in many applications, the width of the PPRs corresponding to $R(\varphi)$ rotations is typically small, often spanning 1-3 modules. As a result, we begin to observe benefits from our algorithm once 10 modules, or 100 logical qubits, are involved. We note that the inter-module logical ZZ and XX measurement must be supported (as is the case for bivariate bicycle codes). 

Our parallelism algorithm proceeds as follows. For each circuit layer, we examine the concurrent R($\varphi$) rotations and compute their total synthesis cost, denoted as $\tau(R(\varphi)_i)$. We then dynamically identify the rotation with the largest remaining synthesis cost and prioritize accelerating that sequence via teleportation, while serializing the others. In effect, we speed up one dominant synthesis sequence at a time. This process continues until all $R(\varphi)$ rotations in the layer are completed, after which the algorithm proceeds to the next layer and repeats for the entire circuit. This process is shown in Figure \ref{fig:teleportation-scheme}, where the injection pivot $\ket{a_0}$ corresponds to the rotation with the largest synthesization cost and is accelerated via teleportation, while another pivot $\ket{b_0}$ associated with a lower cost rotation is serialized. Once the teleportation is completed, the algorithm re-evaluates the remaining rotations, selects the new highest cost rotation, and repeats. 

\begin{figure*}
    \centering
    \begin{subfigure}{0.48\textwidth}
    \centering
    \includegraphics[width=0.95\linewidth]{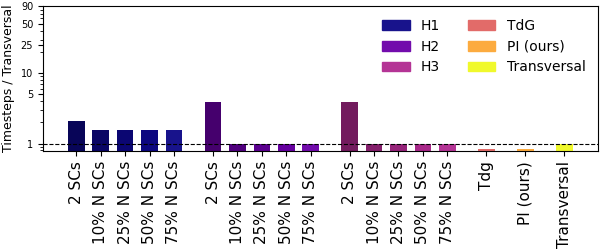}
    \caption{2D Heisenberg Hamiltonian (50 qubits)}\label{fig:QA1}
    \end{subfigure}
    \begin{subfigure}{0.48\textwidth}
    \centering
    \includegraphics[width=0.99\linewidth]{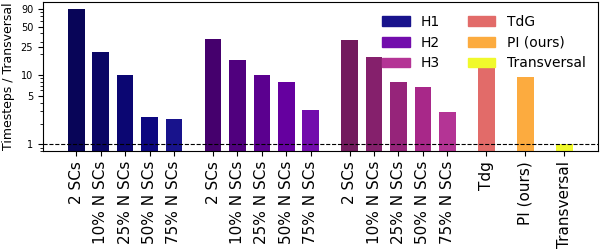}
    \caption{2D long-range transverse field Ising model (100 qubits)}\label{fig:QA2}
    \end{subfigure}
    \begin{subfigure}{0.48\textwidth}
    \centering
    \includegraphics[width=0.99\linewidth]{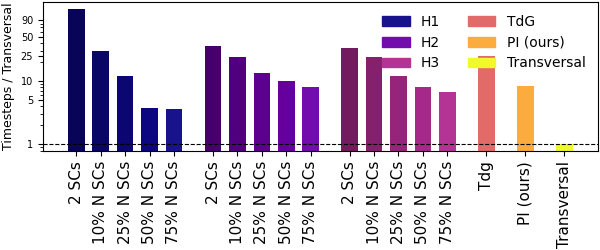}
    \caption{2D nearest neighbor Ising model (100 qubits)}\label{fig:QA3}
    \end{subfigure}
    \begin{subfigure}{0.48\textwidth}
    \centering
    \includegraphics[width=0.99\linewidth]{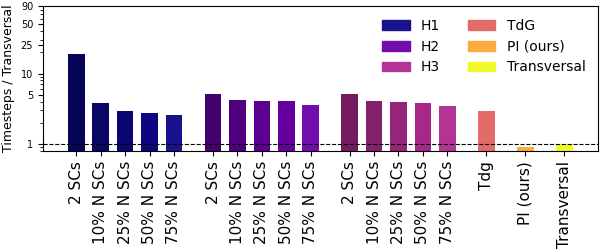}
    \caption{Fermi Hubbard model (200 qubits)}\label{fig:QA4}
    \end{subfigure}
    \caption{Four quantum advantage benchmarks and the relative performance of all three hybrid policies with idealized T state production, the baseline extractor Tour de Gross architecture, and our parallelized injection extractor scheme (Section \ref{sec:parallelize-injection}). The benchmarks shown are: 2D Heisenberg Hamiltonian, 2D long-range transverse field Ising model, 2D nearest neighbor Ising model, and the Fermi-Hubbard model.}
\end{figure*}

In Figure \ref{fig:modules_utilized}, we show that our scheme significantly increases module utilization compared to the base extractor architecture, which leaves many modules idle. Figures \ref{fig:qasmbenchtime}, \ref{fig:qasmbench_spacetime}, \ref{fig:QA1} -- \ref{fig:QA4} and \ref{fig:SQA1} -- \ref{fig:SQA4} demonstrate that our parallelization scheme achieves the best average spacetime cost and outperform all architectures except transversal in total timesteps across QASMBench. 

In Figures \ref{fig:QA1} -- \ref{fig:QA4} specifically, we show time improvements for the 2D long range TFIM, 2D nearest neighbor TFIM, and Fermi-Hubbard applications. Notably, in the Fermi-Hubbard application case with 200 logical qubits, our parallelized extractor architecture surpasses transversal implementation in total timesteps, whereas the base extractor architecture does not. This highlights how the benefits of our parallelization scheme scales with application size. For smaller applications, like the 50 qubit Heisenberg Hamiltonian, parallelization cannot be achieved because there are insufficient unutilized pivots to make up for the preparation and teardown cost. 

We note that this scheme introduces additional inter-module measurement operations, and therefore potentially additional error sources. However, these occur at the scale of R($\varphi$) operations, whereas the speedup is on the order of T-gate injections, making the tradeoff favorable. This can more easily be seen experimentally in the explicit constructions, where Tables \ref{tab:heisenberg} -- \ref{tab:fermi} shows almost no change in error.

\section{Comparing Fault Tolerant Architectures}

\subsection{Assumptions}
We compare the possible QEC architectural frameworks for early fault tolerant demonstrations: single code architectures, extractor only architectures, and hybrid architectures. For this section, we assume that T-state demand can be met such that T-state production is not a time bottleneck for application wall time. This allows for fair comparisons between logical compilation strategies without results being diluted from limited T-state throughput. Later, we revisit more practical scenarios and conduct a detailed study in Section \ref{sec:t-demand}, where we know that our benchmarks require at most $\sim 10^8$ T gates, and show that it is possible to find cultivation factories within the error budget and reasonable spacetime constraints. 

Our analysis initially focuses on two practical and well-studied code families, high-rate QLDPC codes (bivariate bicycle codes) and topological codes that admit transversal gates (surface codes). We use these two codes as representative samples from each of their respective classes, and show that our results can be generalized to any high-rate/topological code (memory/compute) pair that admits QLDPC surgery. For this reason, we study sensitivity to different choices of these pairs and their effect on spacetime tradeoffs  (see Figure \ref{fig:sensitivity-codes}). 

We ensure that the distance of the codes are sufficient to enable successful execution of each benchmark. For surface codes and the two gross code, the distances are at 17 and 18, respectively. While some codes support single shot decoding \cite{single_shot_qec}, which would yield a constant time improvement, we assume $O(d)$ measurements are performed per fault-tolerant instruction across all architectures for fairness and stay consistent with the unit of a timestep equal to $d$ measurements as defined in Section \ref{sec:parallelize-injection}. All of our modeled instructions thus take a cost of one timestep in this section except for qLDPC surgery, which requires 2 timesteps ($2d$ cycles).

\begin{figure}[htbp!]
    \centering
    \includegraphics[width=\linewidth]{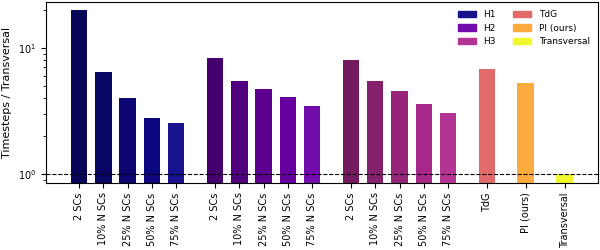}
    \caption{Total \textit{timesteps} required by various hybrid compilers on the QASMBench benchmark suite (H1,H2,H3, as detailed in Section~\ref{sec:hybrid-comparison}). We compare against the base extractor architecture Tour de Gross (TdG) \cite{tourdegross}, and our parallelized injection scheme (Section \ref{sec:parallelize-injection})}
    \label{fig:qasmbenchtime}
\end{figure}

\subsection{Transversal Architectures are Suboptimal in Spacetime}
Transversal implementations allow application level gate operations to be realized directly as logical operations (barring resource states), requiring one $O(d)$ measurement cycle per layer for syndrome extraction. Therefore, the amount of timesteps for an ideal transversal implementation is simply the depth of the synthesized circuit. For codes with high encoding rates, individually addressing logical qubits within a module remains a challenging open problem, particularly when attempting to perform these operations in parallel \cite{qgpu}. Moreover, recent work shows that fully transversal and addressable implementations for a single code with more than one logical qubit is not possible~\cite{no-go-addressable_gottesman}, adding to the challenges of implementing high rate transversal architectures. Some work has investigated performing domain specific subroutines with high-rate transversal codes that avoids requiring a fully addressable and transversal gate set, however such schemes lose generalizability and flexibility \cite{rasql}. For this reason, we use low-encoding rate codes such as the surface code and color code with higher spatial overhead for transversal architectures, meaning that even if transversal architectures are optimal in time, they may be suboptimal in spacetime. 

There are also practical regimes where extractor-based architectures can be time optimal relative to transversal or exhibit similar time performance, even if computing with transversal gate sets on high rate codes becomes a viable path in the future. \textbf{We identify two qualification criteria under which extractor architectures outperform transversal architectures in time:}

\begin{figure}[htbp!]
    \centering
    \includegraphics[width=\linewidth]{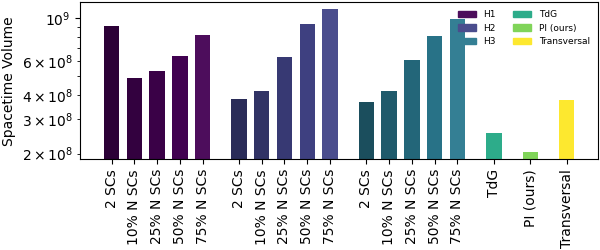}
    \caption{Total \textit{spacetime volume} required by various hybrid compilers on the QASMBench benchmark suite (H1,H2,H3, as detailed in Section~\ref{sec:hybrid-comparison}). We compare against the base extractor architecture Tour de Gross (TdG) \cite{tourdegross}, and our parallelized injection scheme (Section \ref{sec:parallelize-injection})}
    \label{fig:qasmbench_spacetime}
\end{figure}

\begin{enumerate}
\label{sec:rz-criteria}
    \item  \textit{R($\varphi$) synthesis comprises the majority of the circuit}. In some QASMBench benchmarks~\cite{10.1145/3550488}, like the adder and multiplier, R($\varphi$) rotations are trivial and their synthesis cost is negligible. As well, the T ratio (ratio of Ts to R($\varphi$)s), which is a function of the approximation epsilon, influences this cost of how heavy the R($\varphi$) synthesis should be weighted. For most realistic benchmarks, the T ratio is high and varies as a function of $\epsilon$ \cite{ross-selinger}.
    \item  \textit{The transversal circuit cannot parallelize more than $C\%$ of the synthesis cost, where $C$ is the ratio of Cliffords to Non-Cliffords} without $O(C)$ additional space and $O(log(C))$ time to prepare a GHZ state. In the more likely case that the circuit cost is dominated by $R(\varphi)$ rotations (non-Clifford parts, as in most useful FTQC benchmarks), synthesis cost of these R($\varphi$)s becomes the predominant time factor. We find that empirically there is a 60\% reduction in Clifford depth as concurred by \cite{TACO}. Some work has focused on optimizing Clifford depth with the caveat of different resource states and native gates \cite{TACO}. More generally, Clifford commutation brings a reduction on the order of synthesis length while extractor architectures additional instructions are on the order of $R(\varphi)$ rotations, which are outnumbered $1:\epsilon_\text{synth}$.
\end{enumerate} 

\subsection{Hybrid Architectures are Suboptimal in Spacetime}
\label{sec:hybrid-comparison}
Hybrid architectures aim to balance the time advantages of transversal compilation paradigm with the spatial efficiency of a high rate QLDPC code memory. Prior proposals such as \cite{Viszlai_generalized_bb} and \cite{constantoverheadqecneutralatoms} constrain all logical computation to the surface code. However, recent advances in extractor-based computation \cite{he2025extractorsqldpcarchitecturesefficient, tourdegross} remove the necessity of this constraint and expand the design space for hybrid strategies. These newer approaches allow individual modules to be used for compute, enabling compute eviction to proceed while compute in memory proceeds. However, this choice of blocking the entire module (i.e. the $size(M) - 1$ other logical qubits must stall) may not be the globally optimal choice. This further complicates the already large space of choices for different hybrid architecture compute replacement policies/compilation. To evaluate this space, we compare \textit{three} hybrid proposals, varying the number of surface code compute blocks $K$ from the minimum required to admit two-qubit gates ($K=2$) up to $K = 75\% \times N$ where $N$ is the number of logical qubits in the application. Our three hybrid policies are listed as follows, and correspond to Figure \ref{fig:hybrid-archs-explain}(a--c) accordingly:
\begin{enumerate}
    \item \textbf{H1}: Inspired by Refs.~\cite{Viszlai_generalized_bb} and \cite{constantoverheadqecneutralatoms}, all logical computation is mapped to the surface code. The "compute replacement policy" is greedy, using layer lookahead to evict qubits not required in the immediate future. Since all computations are performed in the surface code, T-state injections are also performed there.
    \item \textbf{H2}: Inspired by HetEC \cite{HetEC}, this compiler retains the greedy eviction policy of H1 but allows single qubit Clifford gates to be executed in memory. As observed in Ref.~\cite{HetEC}, in-memory utilization is very low in practice due to limited Clifford commutation. This restricts possible advantages of practical ``dual-computation,'' where in-memory and compute region operations are parallelized. We update HetEC's policy by incorporating the exact cost of each individual Clifford $t_c$ from the bivariate-bicycle ISA defined in \cite{tourdegross} which range from 1--24 timesteps (averaged value of 18.5). We only perform this single qubit Clifford if $t_c < \ell$ where $\ell$ is the load/store cost of four timesteps. While this enables additional parallelism, it stalls all other logical qubits within the same module from further operations for $t_c$ steps. This is a heuristic choice and may not be a globally optimal decision. T-state injections remain restricted to the surface code.
    \item \textbf{H3}: Our proposed hybrid compiler extends H2 by enabling T-state injections to occur directly in memory when beneficial. Specifically, T-state injection is performed only when the T injection cost is $t_i < \ell$, while maintaining the H2 condition $t_c < \ell$ for in-memory single qubit Cliffords. Operation costs are modeled using the same bicycle ISA as in H2. As before, increasing parallelization for the chosen in-memory compute modules comes at the cost of blocking access to all other logical qubits in their respective modules for $t_i$ or $t_c$ steps.
\end{enumerate}

In Figure \ref{fig:qasmbenchtime}, we report the average number of timesteps across a suite of QASMBench Benchmarks \cite{10.1145/3550488}. We observe that hybrid compilation strategies (H1--H3) do not significantly improve performance relative to transversal architecture. At small $K$, hybrid architectures suffer from excessive thrashing operations, yielding time costs comparable to extractor architectures while having additional spatial overhead. As $K$ increases, the relative improvement in time is not proportional to the increase in space.

This tradeoff can be seen more clearly in Figure \ref{fig:qasmbench_spacetime}. At small $K$, hybrid strategies are limited by excessive thrashing, and increasing $K$ increases spacetime cost. Thus, hybrid architectures are suboptimal in spacetime compared to extractor-based architectures. Importantly, spacetime is an optimistic metric for hybrid/transversal approaches because it implicitly weights space and time equally. In practice, particularly for early fault-tolerant neutral atom systems, physical qubit count is the more constrained resource. This amplifies the disadvantage of hybrid architectures.

\begin{figure}[htbp!]
    \centering
    \includegraphics[width=\linewidth]{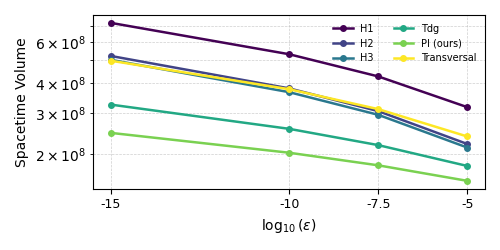}
    \caption{Sensitivity analysis of extractor architectures to synthesis precision $\epsilon$. We evaluate how varying synthesis precision $\epsilon$ affects the spacetime efficiency.}
    \label{fig:sensitivity_epsilon}
\end{figure}

In Figure \ref{fig:sensitivity_epsilon}, we evaluate sensitivity to the synthesis precision $\epsilon$ of $P(\varphi)$ rotations. The spacetime overhead for even an extremely weak approximation of $\epsilon=10^{-5}$ is still worse for hybrid architectures (where we take the best spacetime cost 2 - $75\%N$ for each hybrid policy) when compared to extractor-based architectures. At higher precision $\epsilon = 10^{-15}$, the spacetime gap widens further, explained by the increasingly $R(\varphi)$ serial workloads. 

\begin{figure}[htbp!]
    \centering
    \includegraphics[width=\linewidth]{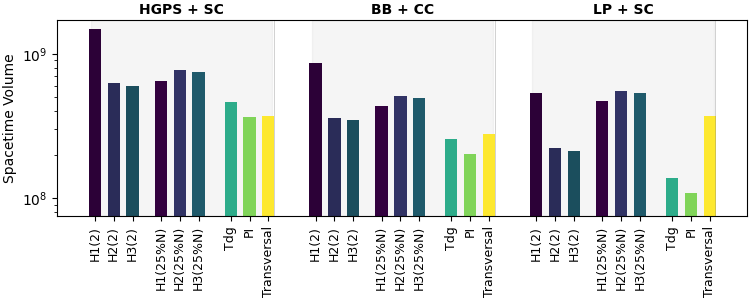}
    \caption{Sensitivity to different hybrid code configurations of high-rate code as memory with a transversal admitting topological code: Hypergraph Product Simplex (HGPS) code + Surface code (SC), Bivariate Bicycle code (BB) + Color code (CC), and Lifted Product (LP) code + Surface code (SC). We report the total spacetime volume for each configuration.}
    \label{fig:sensitivity-codes}
\end{figure}

Additionally, we study sensitivity to different code configurations, showing the generalization of this section's results to alternative memory-compute code pairings in terms of the space time requirements in Figure \ref{fig:sensitivity-codes}. Specifically, we consider two choices for qLDPC codes for memory at similar distance to the two-gross code, a  hypergraph product simplex code ([[1922,50,16]]) and lifted product code ([[1020,136,<20]]). We pair these with topological codes for compute at sufficient distances: a surface code ([[289,1,17]]) or a color code ([[217,1,17]]). All codes were given $n$ ancillary qubits. Because a universal extractor-based instruction set for hypergraph product or lifted product codes is not yet well defined, we are not able to calculate precise utilization of in-memory computation, and assume that the utilization of in-memory logical operations will continue be low, as observed for BB codes. Under this assumption, the spacetime cost is dominated by spatial efficiency. As expected, the lifted product code achieves the best spacetime performance due to its highest encoding rate, while the hypergraph product simplex code performs worst among the memory candidates.

\begin{figure*}
    \centering
    \begin{subfigure}{0.48\textwidth}
    \centering
    \includegraphics[width=0.99\linewidth]{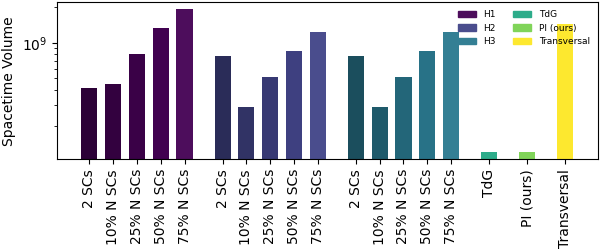}
    \caption{2D Heisenberg Hamiltonian (50 qubits)}\label{fig:SQA1}
    \end{subfigure}
    \begin{subfigure}{0.48\textwidth}
    \centering
    \includegraphics[width=0.99\linewidth]{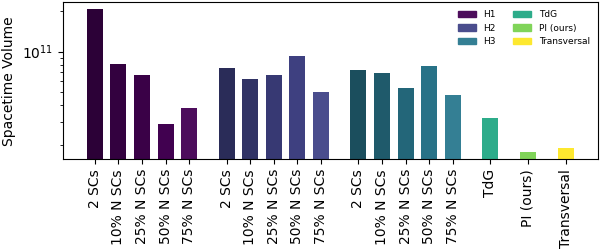}
    \caption{2D long-range transverse field Ising model (100 qubits)}\label{fig:SQA2}
    \end{subfigure}
    \begin{subfigure}{0.48\textwidth}
    \centering
    \includegraphics[width=0.99\linewidth]{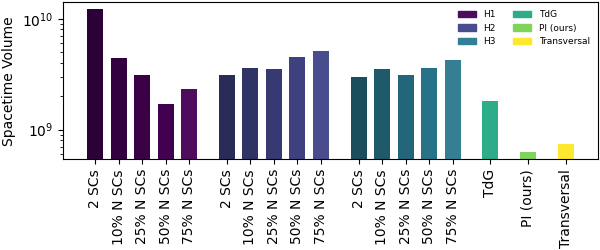}
    \caption{2D nearest neighbor Ising model (100 qubits)}\label{fig:SQA3}
    \end{subfigure}
    \begin{subfigure}{0.48\textwidth}
    \centering
    \includegraphics[width=0.99\linewidth]{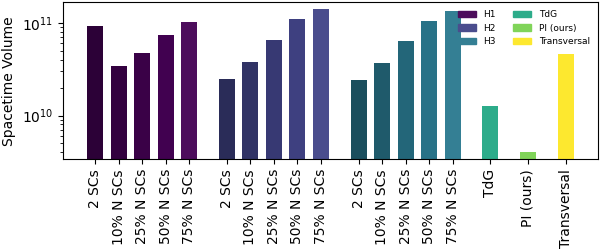}
    \caption{Fermi Hubbard Model (200 qubits)}\label{fig:SQA4}
    \end{subfigure}
    \caption{Four quantum advantage benchmarks and the \textit{spacetime} performance of all three hybrid policies with idealized T state production, the baseline extractor Tour de Gross architecture, and our parallelized injection extractor scheme (Section \ref{sec:parallelize-injection}). The benchmarks shown are: 2D Heisenberg Hamiltonian, 2D long-range transverse field Ising model, 2D nearest neighbor Ising model, and the Fermi-Hubbard model.}
\end{figure*}

\section{Using Extractor Architectures to Demonstrate Early Quantum Advantage}

In previous sections, we developed practical methods for constructing architectures with the lowest spacetime cost, such that we can best find practical early fault tolerant demonstrations. In this section, we build on those results to provide explicit constructions, mainly finding adequate amounts of spacetime-efficient T factories within the error budget, and ensuring that we can actually find good shuttling/gate costs when modeled at the low level. Our two-gross and cultivation pairing \cite{kaavya-cultivation} leads to explicit constructions that we argue as reasonable for early fault tolerant demonstrations. To our knowledge, we demonstrate the first \textit{simulations} of successful execution of quantum advantage applications working within practical resource limits, improving upon the precision of state-of-the-art \textit{resource estimates}~\cite{cain2026shorsalgorithmpossible10000, prakash-beverland-resource-estimates, quantum_algos_table, tourdegross}. 

\subsection{Meeting T Demand}
\label{sec:t-demand}

In previous sections, we assumed an unlimited supply of T-states in order to isolate the architectural effects and enable fair comparisons across schemes. In practice, however, T-state production is a bottleneck. We therefore study how an extractor-based scheme degrades in performance under realistic T-factory constraints. We leverage the recent cultivation protocol from Ref. \cite{kaavya-cultivation} to perform low-overhead T-state production on neutral atoms using native gates, achieving output fidelities on the order of $10^{-10}$. Using the discard parameters reported in that work, we simulate logical compilation under realistic T demand.

\begin{figure*}
    \centering
    \begin{subfigure}{0.48\textwidth}
    \centering
    \includegraphics[width=0.99\linewidth]{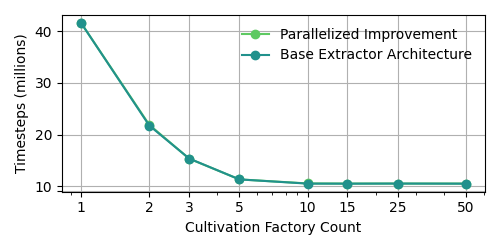}
    \caption{2D Heisenberg Hamiltonian (50 qubits)}\label{fig:QA1_factory}
    \end{subfigure}
    \begin{subfigure}{0.48\textwidth}
    \centering
    \includegraphics[width=0.99\linewidth]{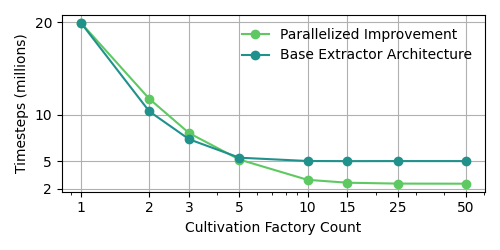}
    \caption{2D long range transverse field Ising model (100 qubits)}\label{fig:QA2_factory}
    \end{subfigure}
    \begin{subfigure}{0.48\textwidth}
    \centering
    \includegraphics[width=0.99\linewidth]{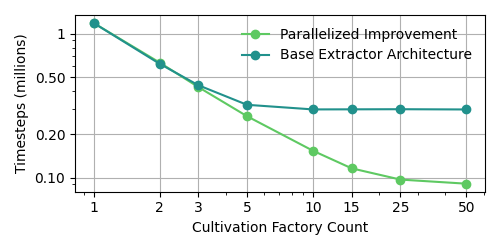}
    \caption{2D nearest neighbor transverse field Ising model (100 qubits)}\label{fig:QA3_factory}
    \end{subfigure}
    \begin{subfigure}{0.48\textwidth}
    \centering
    \includegraphics[width=0.99\linewidth]{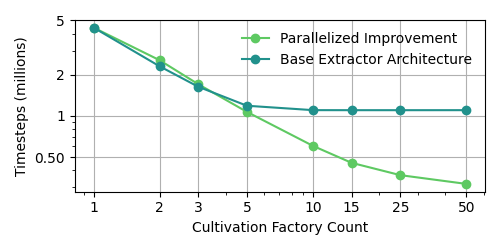}
    \caption{Fermi Hubbard model (200 qubits)}\label{fig:QA4_factory}
    \end{subfigure}
    \caption{Accounting for cultivation T throughput, the performance of our injection scheme and the base tour de gross compilation scheme is shown. The benchmarks are as follows: 2D Heisenberg Hamiltonian, 2D long range transverse field Ising model, 2D nearest neighbor transverse field Ising model, and the Fermi Hubbard model}
\end{figure*}

Figures \ref{fig:QA1_factory}, \ref{fig:QA2_factory}, \ref{fig:QA3_factory}, and \ref{fig:QA4_factory}, show the resulting performance. We observe that performance approaches the ideal regime once roughly 15 cultivation factories are available. In this regime, our scheme achieves up to $\sim 2.5\times$ speedup relative to the base extractor architecture. Of course, our injection protocol is much more dependent on T demand rates as compared to the base extractor architecture. In the regime where both schemes are extremely T-starved, both schemes exhibit similar costs. However, because our scheme incurs a constant preparation and teardown overhead, our scheme can become slightly more costly than the base extractor architecture. Encouragingly, speedups begin to emerge with as few as 3 to 5 factories, corresponding roughly to the regime where ($1 - $discard rate) $\times$ (number of factories) $\approx$ 1, yielding an average production rate of approximately one T state per timestep.

\subsection{Finding Gate and Shuttling Schedules}
\label{sec:shuttling-and-gate}
Thus far, we omitted low-level interaction modeling for extractors (gate times and shuttle times), since measurement latency dominates overall runtime and provides a simple timestep abstraction. We now explicitly model the low-level interactions to validate that our conclusions remain unchanged under realistic physical constraints.

\subsubsection{Generalized Gate and Shuttle Scheduling Algorithm}
Our low-level interactions are modeled using the strategy described in Ref.~\cite{fast_bb}, where fast circuit execution times can be achieved by ensuring that all qubits are placed within a maximum interaction distance. All gates are then executed serially, since each gate creates a Rydberg blockade. In theory, this process could be further parallelized, and we have in fact derived an upper bound on the achievable parallelization. 

\begin{figure}[htbp!]
    \centering
    \includegraphics[width=0.99\linewidth]{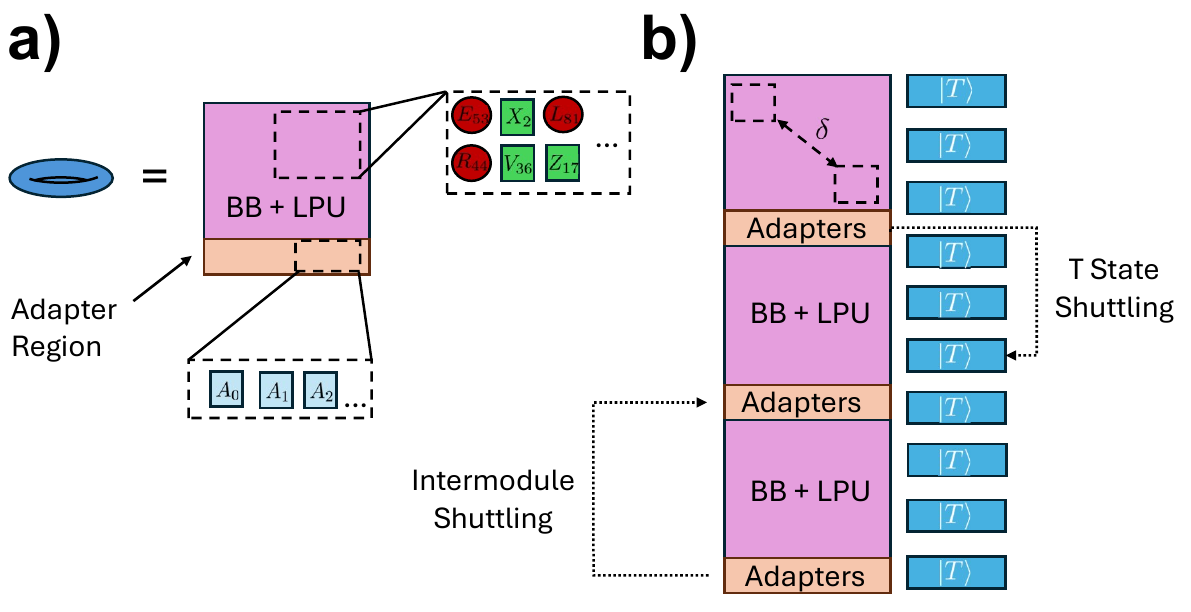}
    \caption{(a) A module is shown decomposed into its lower physical qubits by type. The mixed BB and LPU mapping is shown in pink depicting different left or right data qubits, or X/Z/Vertex/Cycle checks. Adapter checks are shown in the light blue (b) the maximum interaction distance $\delta$ is kept low with the mapping and adapter checks can shuttle in a translationally symmetric fashion such that inter-module measurements can be performed through different modules and T factories.}
    \label{fig:low-level}
\end{figure}

We use simulated annealing combined with the technique described in Ref.~\cite{fast_bb} to get a mapping of the original two gross code and its LPU ancillary system. This relative mapping is then replicated across all modules. The combined mapping of the original code and LPU system is placed above a row of adapter checks, which are used to connect the modules (Figure \ref{fig:low-level}a). All placements corresponding to in-module, idle, and T-factory instructions satisfy the interaction distance threshold, similar to the procedure described in Ref.~\cite{fast_bb}. In our case this max distance was $13.4$ atoms apart for the most demanding instruction. However, inter-module measurements have unbounded interaction distance due to the requirement that adapter checks must be moved to arbitrary target modules. For the adapter-connecting part of the inter-module measurement instruction, we choose to shuttle between adjacent modules, as illustrated in Figure \ref{fig:low-level}b. The total cost in milliseconds of this adapter shuttling takes $0.14M$, where $M$ is the distance measured in module size (24 atoms wide).  

\begin{table*}[t]
\centering
\setlength\tabcolsep{4pt}
\caption{\textbf{2D HEISENBERG HAMILTONIAN (N = 50)} extractor end-to-end simulation results, transversal no-shuttling simulation. Extractor architectures use two-gross codes and transversal architectures use surface codes}\label{tab:heisenberg}
\begin{tabular*}{\textwidth}{@{\extracolsep{\fill}}r|rrr|rrr|rrr}
\toprule
& \multicolumn{3}{c|}{\textbf{Parallelized Injection (ours)}} & \multicolumn{3}{c|}{Base Extractor} & \multicolumn{3}{c}{Transversal} \\
Factories & Qubits & Days & Success (\%) & Qubits & Days & Success (\%) & Qubits & Days & Success (\%) \\
\midrule
1  & 4567  & 92.13 & 88.9 & 4567  & 92.13 & 88.9 & 29687 & 92.13 & 88.9 \\
2  & 5354  & 50.86 & 89.0 & 5354  & 50.45 & 89.0 & 30474 & 53.16 & 88.9 \\
3  & 6141  & 37.49 & 88.9 & 6141  & 37.38 & 89.0 & 31261 & 43.53 & 88.9 \\
5  & 7715  & 28.78 & 88.9 & 7715  & 28.64 & 89.0 & 32835 & 36.25 & 88.9 \\
10 & 11650 & 27.02 & 89.0 & 11650 & 27.06 & 89.0 & 36770 & 31.78 & 88.9 \\
15 & 15585 & 27.04 & 89.0 & 15585 & 27.02 & 89.0 & 40705 & 31.04 & 88.9 \\
25 & 23455 & 27.05 & 89.0 & 23455 & 27.22 & 88.9 & 48575 & 30.84 & 88.9 \\
50 & 43130 & 27.09 & 89.0 & 43130 & 27.06 & 89.0 & 68250 & 30.83 & 88.9 \\
\bottomrule
\end{tabular*}
\end{table*}

\begin{table*}[t]
\centering
\setlength\tabcolsep{4pt}
\caption{\textbf{2D LONG RANGE TRANSVERSE FIELD ISING MODEL (N = 100)} extractor end-to-end simulation results, transversal no-shuttling simulation. Extractor architectures use two-gross codes and transversal architectures use surface codes}\label{tab:longrange}
\begin{tabular*}{\textwidth}{@{\extracolsep{\fill}}r|rrr|rrr|rrr}
\toprule
& \multicolumn{3}{c|}{\textbf{Parallelized Injection (ours)}} & \multicolumn{3}{c|}{Base Extractor} & \multicolumn{3}{c}{Transversal} \\
Factories & Qubits & Days & Success (\%) & Qubits & Days & Success (\%) & Qubits & Days & Success (\%) \\
\midrule
1  & 8347  & 42.96 & 94.6 & 8347  & 42.96 & 94.6 & 58587 & 42.96 & 94.6 \\
2  & 9134  & 26.01 & 94.2 & 9134  & 23.08 & 94.6 & 59374 & 16.77 & 94.6 \\
3  & 9921  & 18.21 & 94.2 & 9921  & 16.76 & 94.6 & 60161 & 11.30 & 94.6 \\
5  & 11495 & 12.21 & 94.2 & 11495 & 12.67 & 94.6 & 61735 & 6.92  & 94.6 \\
10 & 15430 & 7.59  & 94.2 & 15430 & 11.92 & 94.6 & 65670 & 3.63  & 94.6 \\
15 & 19365 & 6.92  & 94.2 & 19365 & 11.92 & 94.6 & 69605 & 2.54  & 94.6 \\
25 & 27235 & 6.72  & 94.2 & 27235 & 11.92 & 94.6 & 77475 & 1.66  & 94.6 \\
50 & 46910 & 6.70  & 94.2 & 46910 & 11.93 & 94.6 & 97150 & 0.97  & 94.6 \\
\bottomrule
\end{tabular*}
\end{table*}

\begin{table*}[t]
\centering
\setlength\tabcolsep{4pt}
\caption{\textbf{2D NEAREST NEIGHBOR TRANSVERSE FIELD ISING MODEL (N = 100)} extractor end-to-end simulation results, transversal no-shuttling simulation. Extractor architectures use two-gross codes and transversal architectures use surface codes}\label{tab:nn}
\begin{tabular*}{\textwidth}{@{\extracolsep{\fill}}r|rrr|rrr|rrr}
\toprule
& \multicolumn{3}{c|}{\textbf{Parallelized Injection (ours)}} & \multicolumn{3}{c|}{Base Extractor} & \multicolumn{3}{c}{Transversal} \\
Factories & Qubits & Days & Success (\%) & Qubits & Days & Success (\%) & Qubits & Days & Success (\%) \\
\midrule
1  & 8347  & 2.53 & 99.7 & 8347  & 2.53 & 99.7 & 58587 & 2.53 & 99.7 \\
2  & 9134  & 1.37 & 99.7 & 9134  & 1.34 & 99.7 & 59374 & 1.00 & 99.7 \\
3  & 9921  & 0.96 & 99.7 & 9921  & 0.97 & 99.7 & 60161 & 0.67 & 99.7 \\
5  & 11495 & 0.61 & 99.7 & 11495 & 0.72 & 99.7 & 61735 & 0.41 & 99.7 \\
10 & 15430 & 0.36 & 99.7 & 15430 & 0.68 & 99.7 & 65670 & 0.21 & 99.7 \\
15 & 19365 & 0.29 & 99.7 & 19365 & 0.68 & 99.7 & 69605 & 0.14 & 99.7 \\
25 & 27235 & 0.25 & 99.7 & 27235 & 0.68 & 99.7 & 77475 & 0.09 & 99.7 \\
50 & 46910 & 0.24 & 99.7 & 46910 & 0.68 & 99.7 & 97150 & 0.05 & 99.6 \\
\bottomrule
\end{tabular*}
\end{table*}

\begin{table*}[t]
\centering
\setlength\tabcolsep{4pt}
\caption{\textbf{FERMI HUBBARD (N = 200)} extractor end-to-end simulation results, transversal no-shuttling simulation. Extractor architectures use two-gross codes and transversal architectures use surface codes}\label{tab:fermi}
\begin{tabular*}{\textwidth}{@{\extracolsep{\fill}}r|rrr|rrr|rrr}
\toprule
& \multicolumn{3}{c|}{\textbf{Parallelized Injection (ours)}} & \multicolumn{3}{c|}{Base Extractor} & \multicolumn{3}{c}{Transversal} \\
Factories & Qubits & Days & Success (\%) & Qubits & Days & Success (\%) & Qubits & Days & Success (\%) \\
\midrule
1  & 15151 & 9.40 & 98.8 & 15151 & 9.40 & 98.8 & 116387 & 9.40 & 98.8 \\
2  & 15938 & 5.49 & 98.7 & 15938 & 4.95 & 98.8 & 117174 & 4.12 & 98.8 \\
3  & 16725 & 3.80 & 98.7 & 16725 & 3.56 & 98.8 & 117961 & 2.95 & 98.8 \\
5  & 18299 & 2.41 & 98.7 & 18299 & 2.65 & 98.8 & 119535 & 2.02 & 98.8 \\
10 & 22234 & 1.41 & 98.7 & 22234 & 2.49 & 98.8 & 123470 & 1.34 & 98.8 \\
15 & 26169 & 1.12 & 98.7 & 26169 & 2.49 & 98.8 & 127405 & 1.13 & 98.8 \\
25 & 34039 & 0.93 & 98.7 & 34039 & 2.49 & 98.8 & 135275 & 0.98 & 98.8 \\
50 & 53714 & 0.83 & 98.7 & 53714 & 2.51 & 98.8 & 154950 & 0.88 & 98.8 \\
\bottomrule
\end{tabular*}
\end{table*}

\subsection{Quantum Advantage Results}
Using the instruction-level logical error rates and operation times from Table \ref{tab:instructions}, together with our parallelized extractor logical compilation policy, we construct spacetime-efficient early fault tolerant architectures capable of executing quantum advantage applications at high success probability with scientific interest. In Tables \ref{tab:heisenberg} -- \ref{tab:fermi}, we observe that the optimal spacetime volume for our scheme is achieved at around $5-10$ cultivation factories, and the total physical qubit counts remain within the $10,000$-$20,000$ range. Notably, even at these practical spatial capacities, our parallelized injection scheme provides greater than $2\times$ speedup without increasing total qubit count, directly translating to improved spacetime efficiency. Compared to the resource-state-excluding experiments (Figures \ref{fig:SQA1} -- \ref{fig:SQA4}), the large spatial costs of cultivation in the limit of many factories allow for transversal implementations to be more competitive in overall spacetime on Ising model problems. Still, these schemes take $7\times$ more qubits to achieve similar spacetime costs to our parallel injection scheme's peak performance, which is much more practical in the early fault-tolerant scenario where qubits are extremely valuable.

\begin{table}[htbp!]
\centering
\renewcommand{\arraystretch}{1.15}
\caption{Fault Tolerant Instruction Error Rates. $M$ indicates the shuttling distance measured in modules spanned.}
\label{tab:instructions}
\begin{tabular}{|l|c|c|}
\hline
\textbf{Instruction} & Error& \textbf{Time} \\
\hline
Idling Operation & $10^{-20}$\cite{tourdegross} & 182 ms \\
Shift Automorphism & $10^{-15}$\cite{tourdegross} & 182 ms \\
In-module Measurement & $10^{-11}$\cite{tourdegross} & 183 ms \\
Inter-module Measurement & $10^{-9}$\cite{tourdegross}& $183 + 0.14M$ ms\\
T Cultivation &  $10^{-8}$ \cite{kaavya-cultivation} & 143 ms \\

\hline
\end{tabular}
\end{table}

These results refine existing hardware-agnostic analytical resource estimates, such as those in Refs.~\cite{cain2026shorsalgorithmpossible10000, prakash-beverland-resource-estimates, quantum_algos_table, tourdegross}, with more precise \textit{simulation} on exact hardware. To our knowledge, this is the first end-to-end simulation of quantum advantage applications executed under realistic hardware constraints. Consequently, our results provide concrete targets for attainable hardware parameters and applications to strive for in early fault tolerant demonstrations. 

For completeness, we include the single factory case, however, no parallelized improvement is applied, since parallelization is impossible with only one available T source. This case is shown as a reference for the strictly space-optimal construction.

\section{Evaluation Parameters and Simulation Methodology}
\label{sec:evaluation}
\subsection{Neutral Atoms Machine Constraints}
For our experiments producing Tables \ref{tab:heisenberg} -- \ref{tab:fermi}, we assume hardware parameters listed in Table \ref{tab:params}, which correspond to instruction time/error in Table \ref{tab:instructions}. The gate and shuttling times account for up to $1.4\%$ of execution time in the worst case. If measurement times were more pessimistic, e.g., 80 ms, then gate and shuttling times would account for less than $0.18\%$ of the total execution time in the most gates and shuttling intensive instructions. Conversely, if this number were optimistic, e.g., 1 ms, then shuttling and gate times could contribute up to $\sim 14\%$ of total measurement time.

\begin{table}[h!]
\centering
\renewcommand{\arraystretch}{1.1}
\caption{Experimental and system parameters assumed in all simulations. Chosen parameters were inspired by Refs.~\cite{dolev_gate_and_shuttle_times, survey_neutral_atoms_timings, fast_bb}}
\label{tab:params}
\begin{tabular}{|l|c|}
\hline
\textbf{Parameter} & \textbf{Value} \\
\hline
Gate Time &  1 $\mu$s\\
Shuttling Speed & 0.5 $\mu$m/$\mu$s \\
Atom Spacing & 1.5 $\mu$m \\
Measurement Time &  10 ms\\
Physical Error &  $10^{-3}$\\
Max Interaction Distance & 15 atoms\\
Coherence Time &  100s\\
\hline
\end{tabular}
\end{table}

\subsection{Error Modeling and Success Probability}
For all of our instructions, we assume that the per physical gate atom loss is the most dominating source of error, even though better single and two qubit gate fidelities have been demonstrated \cite{saffman_neutral_atoms_parameters, neutral_atoms_gate_times_1, dolev_gate_and_shuttle_times}. For this reason, we use both the instruction error rates from Ref.~\cite{tourdegross} and the cultivation error rate from Ref.~\cite{kaavya-cultivation} at physical error rate $10^{-3}$. We define total success probability as the probability that all instructions succeed, as seen in \cite{sethi_optimizing_2026, Viszlai_generalized_bb}:
\[\Large(\prod_{i \in I}(1 - p_i)\Large) \times (1 - \epsilon)^R
\]

Where $p_i$ is the probability of each individual instruction's failure rate in the set of all instructions $I$, and $R$ is the number of $R(\varphi)$ gates in the application.

\subsection{Compilation of Explicit Constructions and T Creation}

For both the base extractor and our parallelized scheme, we compile PBC circuits to BB code instructions for the two gross code using our noise modeled instruction set. For the transversal surface code architecture, we keep the Clifford+T gate implementation. We then simulate execution layer by layer, processing $R(\varphi)$ rotations (T gates in the transversal case) according to the studied parallelization policy, and compute the total amount of error and time taken, subject to the changing T availability. For a given rotation in the extractor scheme, each involved module sees an entangling gate (X basis measurement), a chain of intermodule measurements (though this is parallelized while T injections are happening, and is never greater than the T injection cost for our chosen simulation values), in module measurements (18.5 on average per $R(\varphi)$ \cite{tourdegross}), and a Z basis reset.

For the transversal case, we do not explicitly model shuttling and instead compare to a best-case lower bound on time where shuttling/mappings are not accounted for, though we note this likely has little effect on wall time. For the extractor cases we model mapping, shuttling, and gate schedules precisely for end-to-end simulation. The logical qubits are mapped to each module such that each logical qubit $q$ is equal to $\lceil \frac{q}{|M|} \rceil$, where $|M|$ is the size of each module. The corresponding factories are placed horizontal from each module. A given factory is available with a probability $\pi = 1 - \delta_r$, where $\delta_r$ is the discard rate (we use a $\delta_r = 80\%$ ~\cite{kaavya-cultivation}), and the per cycle T throughput is calculated by taking the cultivation time ratio to the timestep ratio ($d$ cycles). We simulate this nondeterminism: at any given timestep during the T injection process, if no T state is available when required, a stall must be triggered. This adds to the total wall time, and the process of generating valid T states continues. The transversal simulation also accounts for factory nondeterminism in this same way. Upon the end of an instruction at a given layer, we synchronize operations such that we wait for the longest inter-module measurement to finish. In the worst case, this is the intermodule measurement with the shuttling distance from the furthest factory to the furthest module. If shuttling paths are translationally symmetric from factories to modules, we group the shuttling instructions into a singular layer. Otherwise, we add this to the cumulative shuttling cost of the layer. We leave the optimization of nonlocal low-level shuttling paths and parallelization of these operations between different layers for future work. Space for extractors is computed as the sum of the cost of the two gross LPU, adapter qubits, logical data and ancilla qubits, and the space of the cultivation factory from Ref.~\cite{kaavya-cultivation} (787 physical qubits). Space for the distance 17 transversal surface code is computed as the sum of the cost of data and ancilla qubits per logical qubit plus the cultivation factories.

\section{Conclusion and Discussion}
Given the recent rapid progress of the neutral atom platform, here we investigate architectures capable of demonstrating scientifically interesting early fault-tolerant quantum advantage.
We first analyzed the cost of extractor-based architectures and identified the dominant source of time overhead.
We then showed that these architectures can be parallelized on neutral atom systems by leveraging their reconfigurable connectivity and otherwise unutilized modules in the logical circuit. 
We proposed an explicit parallelization algorithm and demonstrated speedups of up to $3\times$ over the baseline extractor architecture, which can be achieved even after modeling resource state production and precise movement patterns.

We then compared leading architectural choices and found that transversal and hybrid architectures are suboptimal in overall space-time performance relative to extractor based architectures. 
For certain application classes, our parallelized extractor-based scheme further improves time performance while preserving spatial efficiency.
Incorporating detailed simulations that account for shuttling times, gate durations, and T-state cultivation factories, we identified concrete architectures capable of enabling practical early demonstrations of fault tolerance.

Taken together, these results position parallelized extractor-based architectures as a compelling path toward early quantum advantage on neutral-atom platforms. 
We provide explicit near-term resource estimates and concrete architectural proposals to guide experimental efforts, and motivate reconsidering fully addressable transversal gate sets in favor of parallelized extractor-based compilation strategies.

\bibliographystyle{ACM-Reference-Format}
\bibliography{references}

\end{document}